\documentclass[preprint]{aastex}
\usepackage{natbib}
\shorttitle{Mass evolution in the NICMOS UDF}
\shortauthors{Gwyn \& Hartwick}

\begin{document}


\title{The Stellar Mass Evolution of Galaxies in the NICMOS Ultra Deep Field}


\author{S. D. J. Gwyn\altaffilmark{1} and F. D. A. Hartwick}
\affil{Department of Physics and Astronomy, University of Victoria, PO Box 3055, STN CSC, Victoria, BC, V8W 3P6, Canada}

\email{gwyn@uvic.ca} 
\altaffiltext{1}{Guest User, Canadian Astronomy Data Centre, which is operated by the Herzberg Institute of Astrophysics, National Research Council of Canada.}



\begin{abstract}
We measure the build-up of the stellar mass of galaxies from $z=6$ to
$z=1$. Using 15 band multicolor imaging data in the NICMOS Ultra Deep
Field we derive photometric redshifts and masses for 796 galaxies down
to $H_{AB}=26.5$.  The derived evolution of the global stellar mass
density of galaxies is consistent with previous star formation rate
density measurements over the observed range of redshifts. Beyond the
observed range, maintaining consistency between the global stellar
mass and the observed star formation rate suggests the epoch of galaxy
formation was $z=16$.

\end{abstract}



\section{INTRODUCTION}

For the last ten years, the Lilly-Madau diagram
\citep{cfrssfr,harry,mad1998} has been central to the discussion of
galaxy evolution.  It shows that the star formation rate density
(SFRD) increases with redshift to around $z\sim1$ and decreases beyond
that.  In the last year, thanks to the Ultra Deep Field (UDF), several
points have been added at the high redshift end of the diagram
\citep{bouwens6,bouwens7,bunkerudf,stancdfs}.

While the Lilly-Madau diagram is a useful tool for studying galaxy
evolution, it is subject to some uncertainties, particularly at the
high redshift end. Typically, high redshift galaxies are selected with
the Lyman break method, the UV luminosity function is computed, and
this is then extrapolated out to the faint end and integrated. UV
light is a good tracer of star formation, so multiplying the total UV
luminosity density by a conversion factor yields the star formation
density.  One source of uncertainty lies in identifying the high
redshift galaxies: the Balmer break may be confused with the Lyman
break, putting spurious galaxies {\em in} to the sample, and galaxies
with heavy extinction may be left {\em out}. Extinction must also be
considered when converting the UV flux into a star formation rate.
Typically, a factor of 5 is used for the extinction correction, but the
exact value is imperfectly known.  For these reasons, it would be
satisfying to have some corroboration of the star formation rate.

The star formation rate density (SFRD) is essentially $dM_\star/dt$ where $M_\star$ is
the baryonic mass in the form of stars, normalized to one cubic
megaparsec.  The integral of $dM_\star/dt$ is just $M_\star(t)$, the
global stellar mass density (GSMD).  Deep infrared images and some
means of determining redshift (either photometrically or
spectroscopically) are required to measure the GSMD.

The NICMOS Ultra Deep Field is ideal for this purpose. It has deep
infrared imaging (26.5 $AB$ magnitude) as well as a wealth of imaging
in other bands with which to compute photometric redshifts. The
wavelength coverage extends from U to K in 15 overlapping bands from a
number of different surveys.  This paper describes measurements of
global stellar mass density out to $z=6$ in the NICMOS Ultra Deep Field.

There are a number of measurements of the global stellar mass density
in the literature.  \citet{rudnick}, working with the HDF-South and
the FIRES K data \citep{labbe}, measured the rest-frame optical
properties of galaxies out to $z=3$ and from their average properties
deduced their masses.  \citet{dickmass} used the HDF-North with NICMOS
J and H observations.  They computed the global luminosity density out
$z=3$, computed the average mass-to-light ratio of galaxies at these
redshifts and so deduced the global mass.  \citet{k20}, on the other
hand computed stellar masses for galaxies individually.  They used the
K20 survey, which consists of UBVRIzJK observations with a limiting
magnitude of $K_{\rm Vega}=20$ and spectroscopic redshifts, out to
$z=2$. \citet{gddsmass} used the GDDS survey \citep{gddsdata}, which
is also based on multi-wavelength data and spectroscopic redshifts but
extends slightly deeper and is specially tuned to the wavelength range
$0.8<z<2$.  \citet{drory2004} used the MUNICS survey \citep{munics1}
which is shallower ($K_{\rm Vega}=19.5$) and extends only out to
$z=1.4$.  The highest redshift survey so far is that of
\citet{drory2005}, who used the GOODS-South data and Fors Deep Field
\citep{fors} to go to $z=5$ and $K_{AB}=25.4$.

In Section \ref{sec:data} we describe how we resampled all the various
imaging data and produced the catalog of photometry and photometric
redshifts.  In Section \ref{sec:masses} we derive masses and mass
functions for the galaxies in the sample and examine the evolution of
the global stellar mass density with time.  In Section
\ref{sec:discussion} we compare our measurements of the global stellar
mass density evolution with the \citet{hartwick2004} model, the
predictions from the star formation rate density, and previous work.

Throughout this paper we use the AB magnitude system \citep{abmag} and
adopt the concordance cosmology of $H_0=70~km s^{-1} Mpc^{-1}$,
$\Omega_m=0.3$, $\Omega_\Lambda=0.7$.

\section{DATA}
\label{sec:data}

 
\subsection{Data sources}

The NICMOS UDF field lies within the GOODS South region of the sky and
has been imaged by a large number of telescopes at a variety
of wavelengths. For this project four sources of imaging data for this field
were considered.

\begin{itemize}

\item Space-based infrared imaging: This data was taken with NICMOS
under the HST Cycle 12 Treasury Program and forms the basis for this
work.  It was taken in two bands, F110W and F160W, which roughly
correspond to $J$ and $H$.\footnote{data available at: 
\url{ftp://archive.stsci.edu/pub/hlsp/udf/acs-wfc/}}

\item Space-based optical imaging: This is the Ultra Deep Field
proper.  The data was taken in four bands: F435W ($B$), F606W
(somewhere between $V$ and $R$), F775W ($I$) and F850LP
($Z$).\footnote{data available at:
\url{ftp://archive.stsci.edu/pub/hlsp/udf/nicmos-treasury/}}

\item Ground-based infrared imaging: These images were taken as part
of the GOODS survey with ISAAC on the VLT \citep{isaac}.  There are
three bands: $JHK'$. Unlike the other data sets, the ISAAC data has
been released as one image per pointing with multiple pointings,
rather than a combined mosaic.\footnote{data available at:
\url{http://www.eso.org/science/eis/old$\underline{~~}$eis/eis$\underline{~~}$rel/goods/goods$\underline{~~}$rel$\underline{~~}$isaac.html}}

\item Ground-based optical imaging: These were taken as part of the
ESO Imaging Survey \citep{arnoutsEIS}. There are 6 bands: $U'UBVRI$
($U'$ is slightly bluer than $U$).\footnote{data available at:
\url{http://www.eso.org/science/eis/old$\underline{~~}$eis/eis$\underline{~~}$rel/goods/goods$\underline{~~}$rel$\underline{~~}$other.html}}

\end{itemize}

All of these data have been released in fully processed form. No
additional processing is necessary.  However, the images in each data
set have different scales and sizes than the images in other data
sets.

\subsection{Resampling the Images}
\label{ssec:resamp}

There are two drivers to the photometry: in order to compute
photometric redshifts, it is necessary to have accurate colors. For
this, fixed circular aperture photometry, preferably through a small
aperture, is sufficient.  The photometry is also used for computing
masses of the galaxies; for this, accurate total magnitudes are
necessary. Total magnitudes are best measured through an adaptive
elliptical aperture \citep{kron}.  If there are to be no systematic
shifts, then these apertures must be matched in all bands.  If the
images are registered, this can be done easily using SExtractor
\citep{hihi} in ``double-image mode''.  Since all the images have
different scales, they must be resampled to put them on the same
astrometric grid.

The resampling was done with {\tt SWarp} \citep{swarp}.
{\tt SWarp}'s main purpose is to combine several (not necessarily
overlapping) images into a single image. It resamples the input images
to put them on a common astrometric grid and scales the flux to
correct for any photometric shifts between the images. It can also be
used to just remap a single image with no flux scaling.

The NICMOS $H$ image was used as the base image to which all the other
images were matched.  The NICMOS $J$ image needed no remapping, as it
was already on the same scale. The optical images from the UDF were
remapped. This changed their scale from 0.03~$''$/pixel to
0.09~$''$/pixel, making them slightly undersampled. The ISAAC images
were {\tt SWarp}ed together so that different pointings in each band
were combined into a single mosaic.  The EIS images show a small but
systematic shift in astrometry with respect to the other data
set. Since this shift is several times smaller than the seeing of the
ground-based EIS images, it was noted, but not corrected. Both sets of
ground-based images are greatly over-sampled by the remapping.

The resampling does not introduce any appreciable shifts in either
position or flux. The {\tt SWarp} documentation claims that with
3-pixel Lanczos interpolation, position is conserved to within one
tenth of a pixel in position and flux is conserved to 0.2\%. We
checked this by comparing the properties of objects in the original
images to the same properties in the resampled images.  Position was
measured by centroid, and flux by magnitude in an 1$''$ diameter
aperture.  No significant shifts in either position or flux were
noted.

\subsection{Photometry}
\label{ssec:phot}

{\tt SExtractor} \citep{hihi} was run in double-image mode on all the
images in all the bands. The NICMOS F160W ($H$-band) image was used as
the reference image: the objects were detected in $H$, and
\citet{kron} style variable elliptical apertures ({\tt
MAG\underline{~~}AUTO} in {\tt SExtractor}) were computed using the
surface brightness profile of the galaxies in the $H$-band.  Identical
apertures were then used to measure photometry in the other images.

Initially, we used the photometric zero-points given by the producers
of each image. However, it soon became apparent that these zero-points
were in slight systematic disagreement with each another.  This was
discovered by converting all the magnitudes for different objects into
fluxes and plotting these fluxes as a function of central wavelength
of the filter in question. There are a total of 15 overlapping
photometric bands covering the wavelength range 3500\AA\ to
21500\AA. Therefore, the spectral energy distributions generated in
this manner are very well defined.  It was found that some bands were
systematically higher relative to neighboring bands. To derive the
values of the photometric offsets, we selected the 39 galaxies in the
NICMOS UDF with published redshifts. We fitted the SED's of these
galaxies with template spectra, using a similar method as the
photometric redshift technique described in the next section, but with
the redshift fixed to the published spectroscopic redshift. Again,
systematic offsets were noted between the templates and the measured
SED's. These offsets were not correlated with wavelength, but were
instead correlated with the source of the images. This indicates that
the problem lies not with the templates, but with the photometry. For
example, the difference in image quality between the low-resolution
ground-based data and high-resolution space-based data will cause
light to be systematically scattered out of an aperture chosen from
the space-based image. Rather than delve into the details of the
origins of the shifts, we took a pragmatic approach and applied the
average shift determined from the 39 galaxies.  The zero-point shifts
determined in this manner are small, typically 0.1 magnitudes.

We investigated the completeness limits in terms of total magnitude
and peak surface brightness of the photometry. This was done by adding
artificial galaxies at random locations in the image and then
re-running {\tt SExtractor} to find what fraction of the galaxies could
be recovered. Rather than use completely artificial galaxies, several
bright, isolated galaxies were identified in the image. Small image
sections (thumbnails) around these galaxies were extracted. The
thumbnails were then modified by:

\begin{itemize}
\item Scaling the flux levels: This consists of multiplying the value of
each pixel by a factor. This changes both the total magnitude and the
peak surface brightness simultaneously.
\item Resampling the images: This means spreading the light from a
galaxy over more pixels, or concentrating it into fewer pixels. It was
done with bi-linear interpolation, rather than the more sophisticated
Lanzcos interpolation used by {\tt SWarp}. This leaves the total
magnitude intact (as there is no change in flux), but changes the peak
surface brightness.
\end{itemize}

By a combination of these two modifications, one can artifically fade
images of galaxies of arbitrary total magnitude and peak surface
brightness. The galaxies were chosen to be sufficiently bright such
that the sky noise from the thumbnail after being faded was negligible
relative to the sky noise from the image section to which the
thumbnail was added. The 90\% completeness limit was found to be
$H_{AB}=26.5$ magnitudes in total magnitude and $\mu_H=25$ magnitudes
per arcsecond squared in surface brightness.

The false positive limit was tested by multiplying the original
$H$-band image by -1, and running {\tt SExtractor} as before.  All
``objects'' detected this way are obviously not real. At
$H_{AB}=26.5$, the false positive rate is just under 8\%.
The final catalog contains 796 galaxies down to
$H_{AB}=26.5$.

\subsection{Photometric Redshifts}
\label{ssec:photz}
Photometric redshifts were calculated for all the objects in the
field. The usual template-fitting, $\chi^2$ minimization method
\citep{ls86a,thesis} was used. The photometric data for each galaxy
are converted into spectral energy distributions (SEDs). The magnitude
in each bandpass is converted to a flux (power per unit bandwidth per
unit aperture area) at the central, or effective, wavelength of the
bandpass. When the flux is plotted against wavelength for each of the
bandpasses, a low resolution spectral energy distribution is created.

A set of template spectra of all Hubble types and redshifts ranging
from $z=0$ to $z=10$ is compiled. The redshifted spectra are reduced
to the passband averaged fluxes at the central wavelengths of the
passbands, in order to compare the template spectra with the SEDs of
the observed galaxies.  The basis of the template set are the
\citet{cww} spectra. These are supplemented with the SB2 and SB3
spectra from \citet{kin96}. Note that this is the same set of spectra
used by \citet{benpap}.

All these spectra have been extrapolated slightly into the UV. For the
purposes of photometric redshifts, it has been found that the exact
nature of this extrapolation is unimportant. The size of the Lyman
break imposed on the spectra of high redshift galaxies by the IGM
makes the exact shape of the underlying galaxy spectrum almost
irrelevant. To account for the effects of the IGM, we use the
prescription of \citet{madau}.  Having only a small number of
templates can cause aliasing in photometric redshifts. Therefore, 10
new templates have been created in between each pair of the six
original templates. There are a total of 51 templates.

The spectral energy distribution derived from the observed magnitudes
of each object is compared to each template spectrum in turn. The best
matching spectrum, and hence the redshift, is determined by minimizing
$\chi^2$ as defined by the following equation:
\begin{equation}
\label{eqn:chimin}
\chi^2(t,z)=\sum_{i=1}^{N_f} {(F_i-\alpha T_i(t,z))^2 \over \sigma^2_{F_i}},
\end{equation}
where 
$t$ is the spectral type,
$z$ is the redshift,
$N_f$ is the number of filters,
$F_i$ and $\sigma_{F_i}$ are respectively the flux and the uncertainty in
the flux in each bandpass of the observed galaxy, 
$T_i$ is the flux in each bandpass of the template being considered,
and $\alpha$ is a normalization factor given by:
\begin{equation}
\label{eqn:alphamin}
\alpha={ {\displaystyle \sum_{i=1}^{N_f} {F_i T_i \over \sigma^2_{F_i} }} \over {\displaystyle \sum _{i=1}^{N_f}{T_i^2 \over \sigma^2_{F_i} }}}.
\end{equation}

\begin{figure}
\plotone{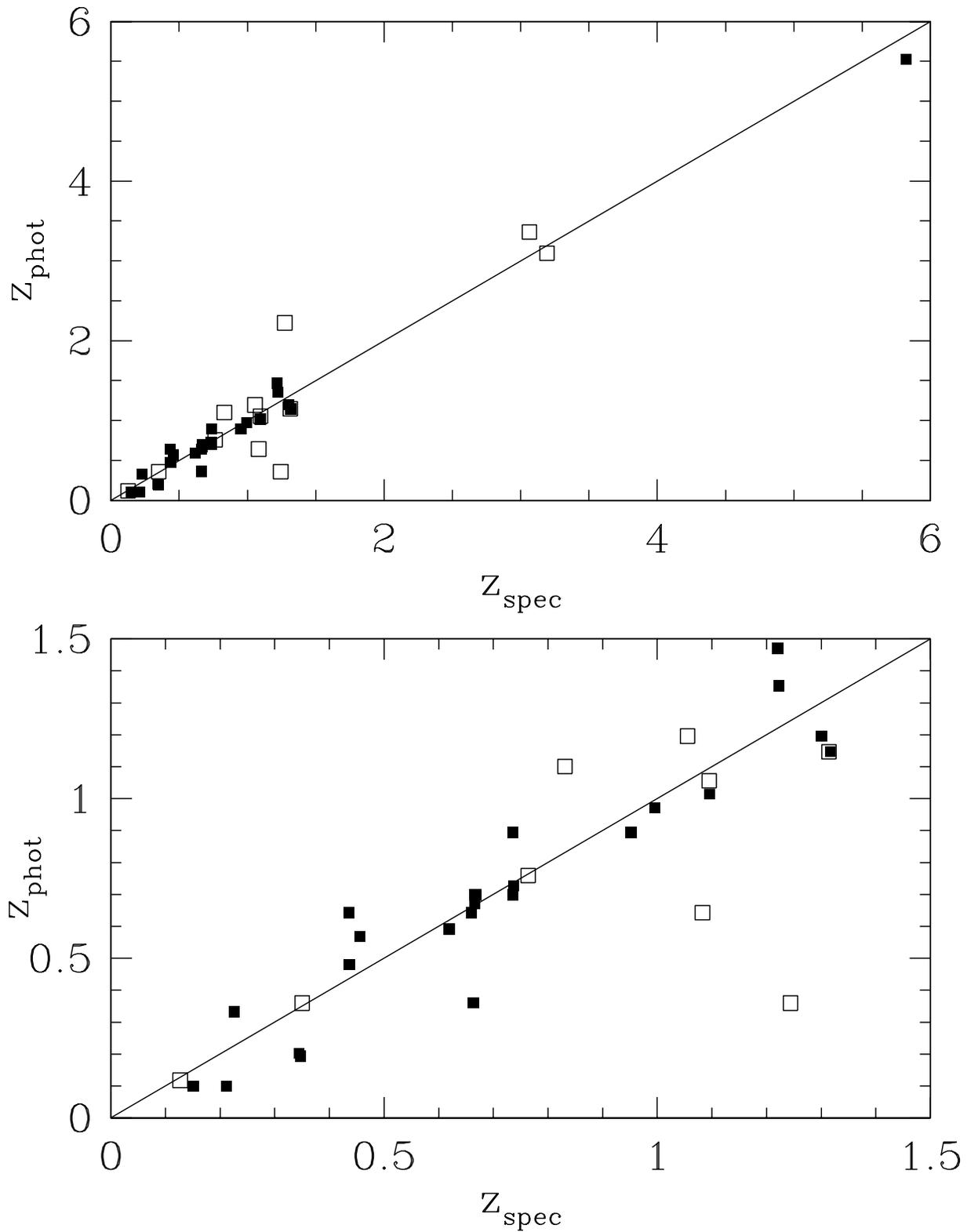}
\caption{Photometric redshifts. The panels show a comparison between
the photometric redshifts and spectroscopic redshifts for our
sample. The upper panel shows the results for all the galaxies with
spectroscopic redshifts in the NICMOS UDF. The lower panel shows the
same thing with an expanded scale for $0<z<1.5$. The filled points
indicate objects with secure redshifts; the open points indicate
doubtful identifications.}
\label{fig:f1}
\end{figure}

Figure \ref{fig:f1} shows a comparison of the photometric redshifts
with the 39 published spectroscopic redshifts.  These were taken from
the work of \citet{vanzcdfs}, \citet{olfcdfs}, \citet{szocdfs},
\citet{croomcdfs}, \citet{stancdfs}, and \citet{strolger} as compiled
by \citet{cdfszlist}. The filled points show objects with secure
redshifts. The open points show objects where the spectroscopic
redshift identification is less secure for one of a number of reasons.
The most common reason is that the ``redshift quality flag'' given in
the relevant paper was low (for example quality=1 or 2 in
\citet{olfcdfs}).  Further, both objects from the \citet{szocdfs}
paper are identified as quasars or AGN.  Since we do not include such
templates in our photometric redshift method, it would be slightly
surprising if we could measure accurate photometric redshifts for
these objects. In fact both objects show good agreement, but are
indicated with open points anyway.  Finally, the position of one of
the spectroscopic redshift objects lies directly between two objects
which are resolved in the HST images. These objects are merged in
lower resolution images.

Taking into account only the objects with secure redshifts, the
photometric redshift error is $\sigma_z = 0.06  (1+z)$ with no
catastrophic failures. Taking into account all the objects increases
this to $\sigma_z = 0.12  (1+z)$, with 2 catastrophic failures
for 39 objects.

The resampled images, and the catalog with photometry and photometric
redshifts are available on the web at:
\url{http://orca.phys.uvic.ca/$\sim$gwyn/MMM/nicmos.html}. Also
available at this site are the {\tt SWarp} and {\tt SExtractor}
configuration files.

\section{MASSES}
\label{sec:masses}

\subsection{Mass Measurements}

We determined masses for each galaxy with a template fitting process
similar to the photometric redshift method described above.  The
photometry for each galaxy is converted into an SED and compared to
series of templates as before.  In this case, the redshift of the
templates being considered is held fixed during the $\chi^2$
minimization.

Rather than the empirical \citet{cww} and \citet{kin96} templates, a
selection of the PEGASE 2.0 galaxy spectral evolution models
\citep{pegase} were used as templates. The templates were redshifted
as before and the IGM correction of \citet{madau} was applied.  Each
of these templates was normalized to $M_\star=1 M_\odot$.
Therefore, the stellar mass of each galaxy is given by:
\begin{equation}
\label{eqn:tmass}
M_\star = \alpha  (4 \pi d_L^2)
\end{equation}
where $\alpha$ is the normalization factor for the best fitting template
described in Equation \ref{eqn:alphamin} and $d_L$ is the luminosity
distance.

The models span the full range of ages from $t=0$ to 14~Gyr. The
metallicity was set to zero (no metals) at $t=0$ in the models. As each
model evolves in time, the metallicity evolves self-consistently.  We
added extinction to the models using the reddening curve of
\citet{calrec}. The amount of extinction was varied from $A(V)=0$ to
1.  The \citet{kroupa} initial mass function (IMF) was used
exclusively.  The template spectra are essentially degenerate with
respect to the IMF.

\subsection{Mass Uncertainties}
\label{ssec:masserr}

While the template fitting process will select one template as being
the best fit, there are several templates that will be consistent with
a given set of photometry and its associated uncertainty.  Using
Equation \ref{eqn:tmass} with these alternate templates gives slightly
different masses. The range of template masses that are still
consistent with the photometry gives an estimate on the random
uncertainty on the mass.  Here, we take all templates which match the
photometry such that $\chi^2 <\chi^2_{best}+1$ (where $\chi^2_{best}$
is the reduced $\chi^2$ of the best fitting template) to be
``consistent''.  The mean mass uncertainty was found to be 0.15
dex. Mass uncertainty, not surprisingly, was found to be an increasing
function of apparent $H$ magnitude.  We adopted $\sigma_{\rm
mass}=0.1$ for $H_{AB}<25$ and $\sigma_{\rm mass}=0.2$ for
$H_{AB}>25$.

A potential source of systematic error is the choice of initial mass
function (IMF).  We chose the \citet{kroupa} IMF rather than the
\citet{salpeter} IMF that is commonly adopted in galactic evolution
studies.  Switching to a Salpeter IMF causes a systematic shift in the
masses of -0.12 dex, in the sense $M_{\rm Kroupa}-M_{\rm
Salpeter}=-0.12$ dex. Switching to the \citet{bg03} IMF causes also
causes a systematic shift of 0.12 dex, but in the other direction.

There are also uncertainties in the mass which are due to the
uncertainties in the redshift. A photometric redshift that is higher
than the true redshift will cause the measured mass to be artificially
higher in addition to any random uncertainties from the template fitting.
We deal with these coupled uncertainties in the following section on mass
functions.

\subsection{Mass Functions and Total Masses}

\begin{figure}
\plotone{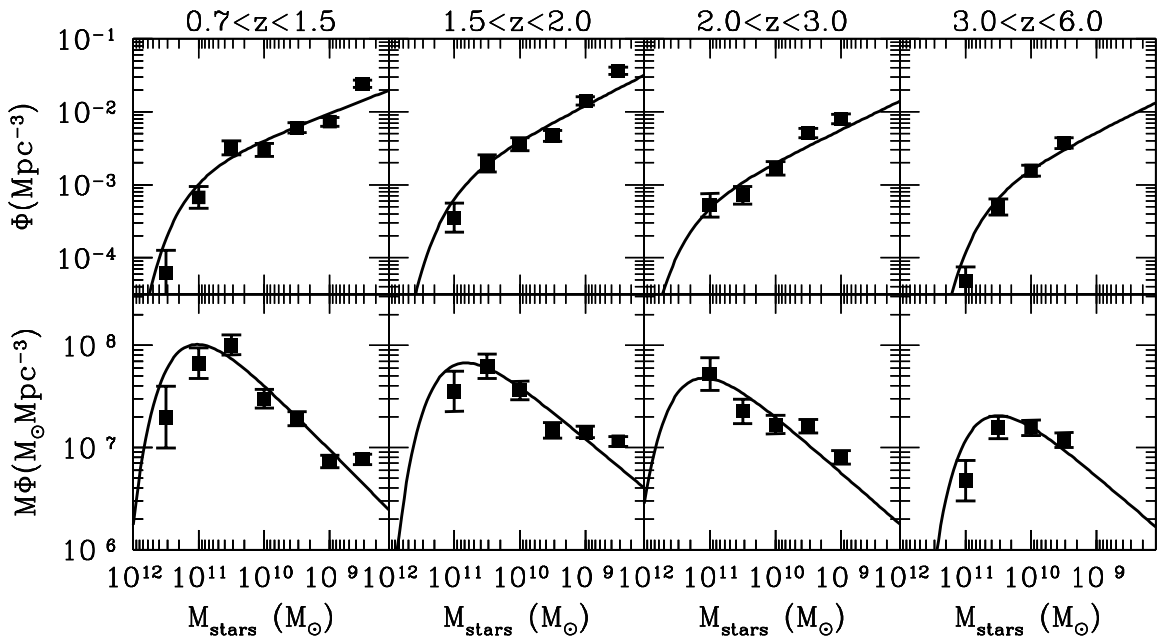}
\caption{Mass functions. The top four panels show the mass functions
for the indicated redshift ranges. The bottom four panels
show the mass functions from the top panels multiplied by 
mass to give the mass density functions.
The lines are \citet{scheck} function fits to the data. 
\citet{edd} corrections
have been applied to the data.
The errorbars indicate Poisson errors only.
}
\label{fig:f2}
\end{figure}

We used the $1/V_a$ method to compute mass functions in a series of
redshift bins.  For each galaxy we measure the accessible volume,
$V_a$, which is the volume the galaxy {\em could} be in and still be
visible in the survey. The sides of this volume are defined by the
edges of the image. The near face of this volume is the lower end of
the redshift bin in question.  The far face of the volume is either
the upper end of the redshift bin or $z_{lim}$, the redshift at which
the object would be fainter than the limiting magnitude of the sample,
whichever is smaller. The limiting redshift, $z_{lim}$, is found by
artificially moving the object out in redshift, and recalculating its
apparent magnitude, taking into account the change in distance modulus
and the $k$-corrections, until it reaches the magnitude limit of the
survey, in this case $H_{AB}=26.5$. The $k$-corrections were computed
by interpolation using the best-fitting template from the photometric
redshift procedure.  The next step is to bin the galaxies by mass and
sum over the galaxies in each bin, weighting by $1/V_a$.  The mass
functions are shown in Figure \ref{fig:f2}.

The upper row of panels in Figure 2 shows the mass functions,
$\Phi(M)$, for four redshift slices. The bottom row of panels shows
the mass density function, $M\Phi(M)$.  Integrating over $M\Phi(M)$
gives the total stellar mass density. The bottom row of panels show that while
we have fairly good coverage on the parts of the mass function that
contribute most to the total mass, we do not cover the full range
necessary. To extrapolate beyond the observed range, we fit
\citet{scheck} functions to the data points:

\begin{equation}
\label{eq:sheck}
\Phi(M)dM=\Phi^\ast (M/M^\ast)^\alpha \exp (-M/M^\ast) d(M/M^\ast),
\end{equation}
where $\Phi^\ast$ is the overall normalization, $\alpha$ is the faint
end slope, and $M^\ast$ is the characteristic mass.  These fits are
shown by the lines in Figure \ref{fig:f2}. The fits are well constrained
around $M^\ast$, well constrained at lower masses (with the possible
exception of the $3<z<6$ redshift bin), and mostly well constrained
at the high mass end, with the notable exception of the $2<z<3$
redshift bin.

\begin{figure}
\plotone{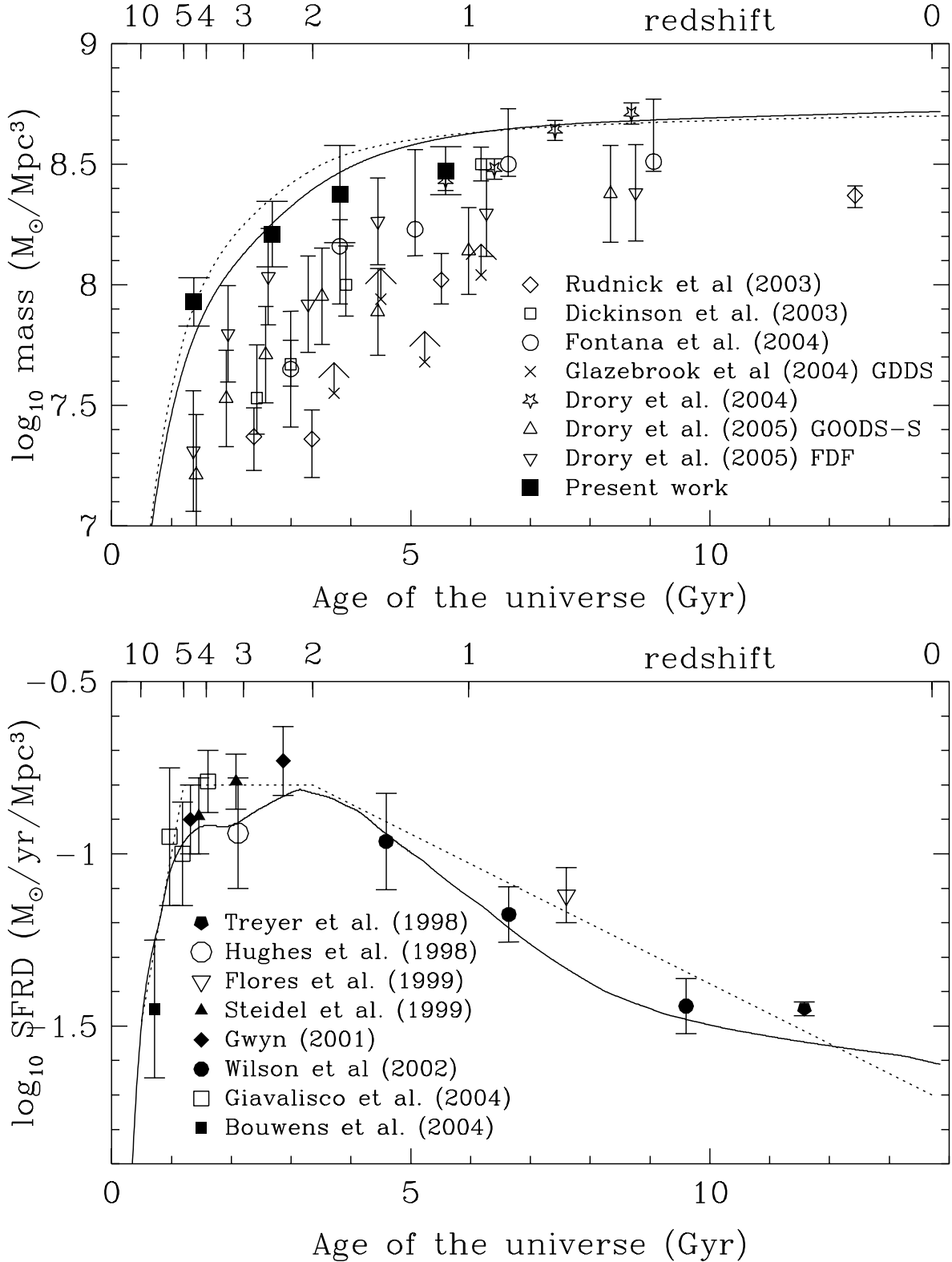}
\caption{Stellar mass evolution. The bottom panel shows the star
formation rate history of the universe. The points indicate
measurements of the star formation rate density from the literature.
The top panel shows the stellar mass evolution from this work along
with data from the literature. The GDDS points are shown as lower
limits.  In both panels, the solid line shows the model of
\citet{hartwick2004}.  The dotted line is a crude fit to the SFR
history (not a model) in the lower panel, and the integral of this fit
in the upper panel.  See main text for references.  }
\label{fig:f3}
\end{figure}

To obtain the total mass density, one must integrate the mass density
function.  This has the analytic form:
\begin{equation}
\label{eq:sheckint}
M_{tot}=\int_0^\infty M \Phi(M) dM = M^\ast\Phi^\ast\Gamma(\alpha+2),
\end{equation}
where $\Gamma$ is the Gamma function.
The results of this integration for the four redshift bins are
shown as solid points in the top panel of Figure \ref{fig:f3}.

\subsection{Total Mass Uncertainties}

Before turning to a discussion of Figure \ref{fig:f3} it is
necessary to describe the possible uncertainties associated with our
measurements.  Besides the usual Poisson noise, there are three
sources of uncertainty. Instead of spectroscopic redshifts, we use
photometric redshifts which have significantly larger associated
errors. This can shift galaxies out of the correct redshift bin,
affect their $V_a$ weighting, and change the measured mass of the
galaxy. Next, there is some uncertainty associated with the mass
measurement itself, as discussed in Section \ref{ssec:masserr}.
Finally, the coverage of the mass density function is not perfect,
which means that the Schechter fitting described in the previous
section may be insufficiently constrained. 

Furthermore, when one measures the distribution of a parameter with a
non-negligible associated error, systematic effects will be noted, as
first discussed by \citet{edd}. Even if the error bars are symmetric,
more low mass objects will scatter into the high mass bins than high
mass objects will scatter into the low mass bins, simply because there
are more low mass objects to be scattered.  This systematic effect can
be corrected if one has a good understanding of the error.

To investigate the uncertainties and to make the Eddington
corrections, we used a Monte Carlo technique.  To the original catalog
of galaxies we added noise to the measured redshifts
($\sigma_z=0.06(1+z)$) and masses ($\sigma_{\rm mass}=0.1$ for
$H_{AB}<25$ and $\sigma_{\rm mass}=0.2$ for $H_{AB}>25$).  Further, we
simulated the effects of redshift error on the derived masses by
noting the relative shift in luminosity distance caused by the
redshift error, and applying the same shift to the mass. From this
``noisy'' catalog we derived mass function, integrated over fitted
Schechter functions and computed total stellar masses.  The RMS of the
range of total stellar masses derived after 100 actualizations was
used for the error bars in Figure \ref{fig:f3}.  The Eddington
corrections thus calculated were applied to the points.  They were
found to be negligible except for the highest mass bin, where they
were on the order of 0.2 dex. The Eddington corrections adjust the
final masses by about 15\%.

\section{DISCUSSION}
\label{sec:discussion}

\subsection{The Evolution of the GSMD}
Figure \ref{fig:f3} shows the build up of stars in galaxies as
function of time.  The top panel shows the global stellar mass
density, while the bottom panel shows the star formation rate.  The
top panel can be thought of as $M_\star(t)$, while the bottom panel
can be thought of as its derivative, $dM_\star(t)/dt$.  This panel is
the reverse of the Lilly-Madau diagram, with cosmic time instead of
redshift on the horizontal axis.  The bottom panel shows a number of
measurements of the star formation rate from the literature
\citep{treyer1998,hughsubmm,thesis,flores,stei99,wilson,giavalisco2004}.
All the star formation rate data shown in Figure \ref{fig:f3} have
been corrected for extinction by the authors of the individual
papers. The only exception is the \citet{bouwens7} data which
necessitated an extinction correction of 0.8 dex in order to bring it
onto the same system as the other data.  Where appropriate, we have
also applied incompleteness corrections assuming a Schechter
extrapolation with $\alpha=-1.5$. For example, \citet{bouwens7} only
compute the UV luminosity function down to $L=0.3L_\star$.
Extrapolating over the full range of $L$ implies a correction of 0.36
dex.

The upper panel shows our measurements of the global mass density as
solid points, together with a number of measurements from the
literature \citep{rudnick,dickmass,k20,drory2004,drory2005} as
assorted open points.  The authors of the above works have corrected
their data for the unobserved portion of the mass function.  In their
analysis of the Gemini Deep Deep Survey, \citet{gddsmass} have chosen not
to extrapolate beyond what they observe. The GDDS data for galaxies
more massive than $\log(M/M_\odot) >10.2$ are plotted as lower limits.
The corrections for the choice of IMF, discussed in section
\ref{ssec:masserr}, have been applied.

The solid line in both panels of Figure \ref{fig:f3} come from
\citet{hartwick2004}, who derived the global star formation history
from observations of the local universe.  Briefly, the model uses the
distribution in metallicity of stars to derive $dM_\star/dZ$ (where $Z$
is the metallicity) and the age-metallicity relationship for globular
clusters to derive $dZ/dt$ (where $t$ is the age of the universe).
Combining $dM_\star/dZ$ and $dZ/dt$, one obtains $dM_\star/dt \equiv$
SFR, the star formation rate. This simple model does a very good job
of explaining the star formation history of the universe, as shown by
the agreement between it and the observations in the lower
panel. Using this star formation history as an input to the PEGASE 2.0
software, we compute a model global stellar mass density.  The result
is plotted in the upper panel. It is in excellent agreement with our
global stellar mass density measurements.  Note that this agreement is
not dependent on the details of the Hartwick model. Almost
any description of the star formation history which agrees with the
observed star formation rates will, once integrated, produce good
agreement with global stellar mass density. This is illustrated by the
dotted line in Figure \ref{fig:f3}.  On the lower panel, this shows a
crude, three-segment ``connect-the-dots'' description of the star
formation history. In the upper panel the dotted line shows the
results of integrating (again with PEGASE 2.0) this star formation
history.  Again, there is good agreement with our global stellar mass
density calculations.

Note that converting a star formation rate to GSMD is a fairly robust
procedure. It is fairly insensitive to the details of the population
modelling. Indeed, there is only a slight loss of accuracy even if one
makes the extreme assumption that all stars, once created, never die.
This is because (for all reasonable initial mass functions) the bulk
of the mass in stars comes from stars of less than 1 $M_\odot$, which
have a main-sequence lifetime of approximately a Hubble time.  In this
simplified case one can do a straight integration of the $dM_\star/dt$
(with $M_\star=0$ at $t=0$) to determine $M(t)$.

\subsection{Comparison with Previous Work}

There are two things to note in the top panel of Figure
\ref{fig:f3}.  The first is the excellent agreement between our
measurements of global stellar mass density and the integral of the
star formation rate. The other is the rather poor agreement between
our measurements and those of previous researchers. The error bars of
\citet{drory2005} overlap with ours, but on the whole there
remains a systematic offset.  We can give no explanation for the
disagreement between our work and that of other authors.  The offset
is not entirely due to our choice of initial mass function. Even
if the IMF offsets discussed in section \ref{ssec:masserr} are
not applied there is still a significant difference.

If the other authors are correct, then we need an explanation for the
difference between their measurements of the GSMD and the integral of
the SFRD. A number of possibilities have been suggested. For example,
one explanation is a change in the average extinction in galaxies over
time; {\it e.g.}, galaxies have lower extinction at high redshift than
is currently thought.  If this were the case, the extinction
correction would be smaller, and the deduced star formation rate would
be lower at high redshift. Less star formation at high redshift means
less mass build-up.  Alternatively, the initial mass function might
vary with redshift in such a way as to reconcile the GSMD and the
SFRD.  However, if our measurements are correct, no explanation is
required.

\subsection{The Epoch of Galaxy Formation}

From Figure \ref{fig:f3}, we can set limits on the amount of
star formation that went on before $z=4.5$, our highest redshift
point. For the concordance cosmology, $z=4.5$ corresponds to a cosmic
age of $t=1.37$ Gyr. We find that the mass in stars at this point was
$8.63\times 10^7 M_\odot Mpc^{-3}$.  If the epoch of galaxy formation
lies at $t=0$, then the average star formation rate density at $z<4.5$
must have been close to $8.63\times 10^7 M_\odot Mpc^{-3} / 1.37 Gyr =
0.063 M_\odot yr^{-1} Mpc^{-3}$. This is in good agreement with the
results of \citet{bouwens7}. If, on the other hand, the epoch of galaxy
formation was at $z=10$ or $z=6$, the same mass of stars must have been
formed in only 0.9 Gyr or 0.4 Gyr respectively.  Then the
corresponding star formation rates must be $0.097 M_\odot yr^{-1}
Mpc^{-3}$ or $0.21 M_\odot yr^{-1} Mpc^{-3}$.  The $z=6$ figure
represents a very significant star formation rate density, greater
than the currently measured peak of the SFRD at $z=2.5$, and
completely incompatible with the measurements of
\citet{bouwens7}. Turning the problem around, if we assume the {\em
maximal} star formation rate allowed by the error bars on
\citet{bouwens7} at $z=7$, and the {\em minimal} mass allowed by our
highest redshift data point, we compute the {\em latest} possible
epoch of galaxy formation to be $z=16$.

\section{SUMMARY}

We have measured the build-up of the stellar mass of galaxies from
$z\sim6$ to $z\sim1$.  Our measurements are consistent with the
predictions from star formation rate density. The derived evolution
of the global stellar mass density of galaxies is consistent with
previous star formation rate density measurements over the observed
range of redshifts. Our measurements of the global stellar mass
density show an offset with respect to previous measurements of the
GSMD.  If we are to maintain consistency between the global stellar
mass and the observed star formation rate, the epoch of galaxy
formation must be at least $z=16$.

\acknowledgments

\acknowledgments 
F.D.A.H. gratefully acknowledges financial support
from a discovery grant from NSERC .  S.D.J.G was supported partially
from the above grant and from an NSERC CRO grant which supports
Canadian participation in the CFHT Legacy Survey.

\clearpage

\end{document}